\documentclass[a4paper]{jpconf}
\usepackage{graphicx}
\usepackage{color, citesort}
\begin{document}
\title{Interstellar Medium Mitigation Techniques in Pulsar Timing Arrays}

\author{L Levin}

\address{Department of Physics \& Astronomy, West Virginia University, Morgantown, WV 26506, USA}

\ead{Lina.Levin@mail.wvu.edu}

\begin{abstract}
Pulsar Timing Arrays use a set of millisecond pulsars in an attempt to directly detect nanohertz gravitational waves. For this purpose, high precision timing of the pulsars is essential and ultimately a precision of the order of $\sim$100\,ns is required. 
Propagation effects in the interstellar medium cause the radio emission from a pulsar to be dispersed and scattered, introducing time variable delays of the pulses on their way to Earth. If these delays are not properly corrected for, they may cause significant errors in the timing analysis of a pulsar. These proceedings will review the effects of the interstellar medium on pulse arrival times and present some of the techniques used to mitigate the associated time delays from the pulsar signal. Correcting for these delays is essential to providing a higher timing precision and hence to increasing the array's sensitivity to gravitational waves. 
\end{abstract}

\section{Introduction}
Pulsar Timing Arrays (PTAs) are observing a set of high-precision timing millisecond pulsars (MSPs) with the aim to detect nano-Hz gravitational waves (GWs). 
MSPs are extremely stably rotating neutron stars, with times of arrival (TOAs) of the pulses measurable to a few\,$\mu$s precision. 
In the timing procedure of a pulsar, residuals are created by making a timing model based on the parameters of the pulsar, and accounting for each rotation of the star. 
By observing MSPs for a long time span, the timing model is improved and the scatter of the residuals can go down to an RMS of $\sim$100\,ns. To search for GWs, PTAs look for correlations in the timing residuals of several MSPs. Since PTAs are sensitive to nanohertz frequency GWs, which make up a part of frequency space not covered by ground based and space based GW detectors, the efforts of various GW projects are complementary. The main GW sources expected in the PTA frequency band are super massive black hole binaries and cosmic strings, and searches are being carried out both for single sources and for a GW background. 
Currently, three collaborations are undertaking these efforts, namely the European Pulsar Timing Array\footnote{http://www.epta.eu.org} (EPTA; \cite{kra13}), the Parkes Pulsar Timing Array\footnote{http://www.atnf.csiro.au/research/pulsar/ppta} (PPTA; \cite{hob13}), and the North American Nanohertz Observatory for Gravitational Waves\footnote{http://nanograv.org} (NANOGrav; \cite{mcl13}). Together they make up the International Pulsar Timing Array\footnote{http://www.ipta4gw.org} (IPTA; \cite{man13}). 

To succeed in detecting GWs, it is necessary to achieve as high timing precision as possible, over a long time span. 
As the signal from a pulsar travels through the ionized plasma of the interstellar medium (ISM), the pulses get scattered and the TOAs of the pulses are delayed. The two dominating ISM effects are dispersion and scattering, and both these effects introduce perturbations to the TOAs and will increase the timing error if not properly corrected for. 
Monitoring of ISM parameters and development of correction techniques is on-going within several pulsar groups \cite{cor10,arc14,lev14}, and ISM mitigation is increasingly recognized as one of the most important parts of the noise budget in PTAs \cite{sti13,kei13,lee14}. 

This document is written as part of the 10th International LISA Symposium conference proceedings, and is based on a talk reviewing the challenges of ISM propagation effects on pulsar timing data as well as the different techniques used to mitigate these effects from the pulsar signals.

\section{Dispersion}
\begin{figure}
	\begin{center}
		\includegraphics[height=0.35\textwidth]{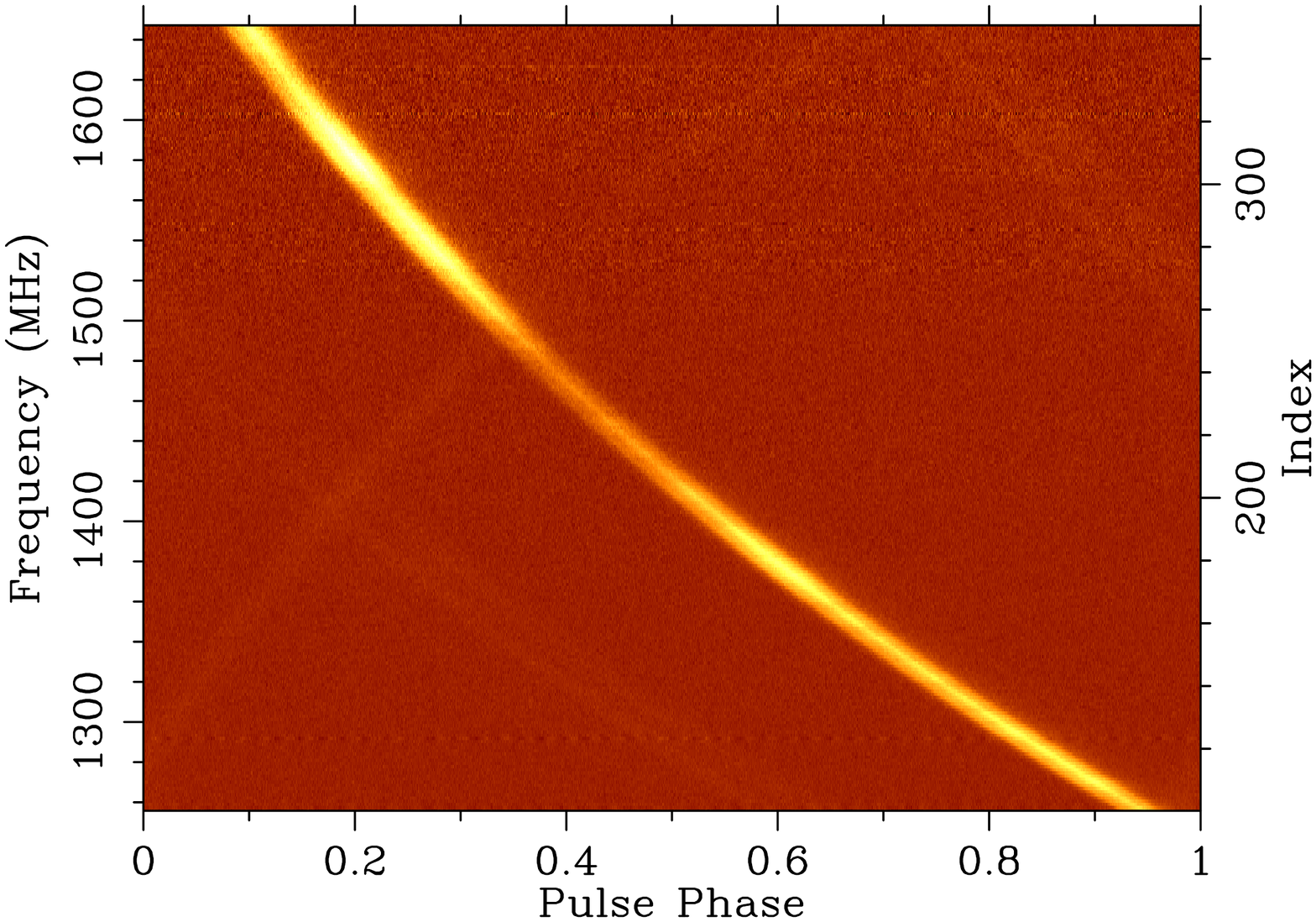}
		\includegraphics[height=0.35\textwidth]{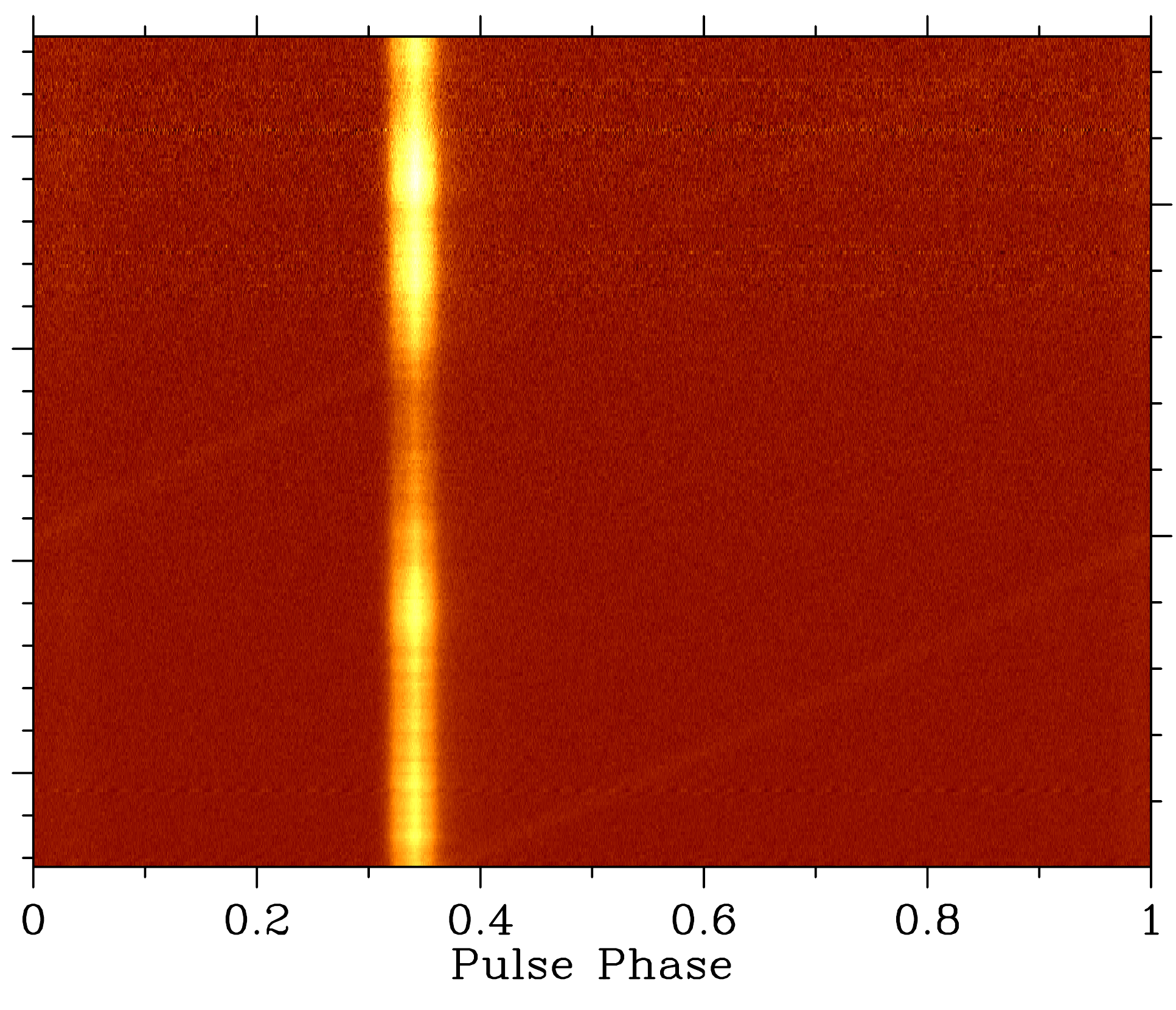}
	\caption{Wide-band observation of PSR\,J1744--1134, before ({\it left}) and after ({\it right}) DM correction. In this case, the dispersion delay over a 400\,MHz band is almost equal to one full rotation of the pulsar. Due to scintillation, some intensity modulation over the band is also visible in this observation. These plots were created from raw data, contributing to the weak interference signals visible in the background.}
	\label{fig:dm}
	\end{center}
\end{figure}

As the pulsar signal travels through the ISM, the overall column density of free electrons has dispersive effects on the broadband pulses. We can write the group velocity of the pulse train as 
\begin{equation}
v_{\rm g} = c \sqrt{1 - \frac{\nu_{\rm p}^2}{\nu^2} }
\end{equation}
where $c$ is the speed of light, $\nu$ is the observing frequency and $\nu_{\rm p}$ is the plasma frequency given by 
\begin{equation}
v_{\rm p} = \sqrt{ \frac{n_{\rm e} e^2 }{\pi m_{\rm e}} } \approx 10^4{\rm Hz} \sqrt{\frac{n_{\rm e}}{1 {\rm cm}^{-3}}}.
\end{equation}
Here, $e$ is the electron charge, $m_{\rm e}$ is the electron mass and $n_{\rm e}$ is the electron density in the relevant direction in the sky, all given in cgs units. If looking into the Galactic plane, the value of $n_{\rm e} \approx 0.03$\,cm$^{-3}$, and it gets lower at higher Galactic latitudes. The definition of the Dispersion Measure (DM) as the integrated column density of free electrons along some line of sight can be written as
\begin{equation}
{\rm DM} = \int^{d}_{0} n_{\rm e} \, dl,
\end{equation}
where $d$ is the distance to the pulsar. From this, we can derive the time delay of the radio wave due to the DM as
\begin{equation}
t_{\rm DM}(\nu) = \frac{e^2}{2 \pi m_{\rm e} c\, \nu^2}{\rm DM} \approx 4.15\,{\rm ms}\, \frac{{\rm DM}}{\nu^2_{\rm GHz}}.
\end{equation}
The observational consequence of the dispersive delay is that a pulsar observed with a broad frequency band will show pulses that arrive later at lower frequency than at higher frequency, following the $\nu^{-2}$ law. By observing pulsars at a wide enough frequency band, or at two or more separate frequency bands, the DM can be measured at each observing epoch. An example of an observation that has not been DM corrected can be found in Fig. \ref{fig:dm}. 

One practical use of the DM is as a tool to infer the distance to the pulsar. This can be done with the use of models of the free electrons in the Galaxy. Since there are more free electrons in certain directions in the Galaxy, such as along the Galactic plane, a higher DM does not necessarily mean a larger distance, but also the sky position of a pulsar need to be considered. Currently, the NE2001 model \cite{cor02} is the most comprehensive and widely used model.  

Since the pulsar and the ISM have different relative velocities, the DM of a pulsar is not constant in time, and the variations of the DM need to be tracked and corrected for in the data. This can be achieved by observing pulsars at two or more separate frequencies at a maximum of a few days apart. 
Different methods on how to optimally correct for the DM variations are being used by the different PTAs. 
Currently, the method used by NANOGrav involves fitting the multifrequency data for a DM within a $\sim$10 day window, in a simultaneous fit with all pulsar parameters \cite{dem13}. 
The PPTA technique is very similar to this, but removes the dispersive delay from the timing residuals after the timing solution has been subtracted. This method includes a frequency independent ``common mode'' delay \cite{kei13}, that is introduced to make sure that no GW signal is absorbed into the DM correction. 
Finally, the EPTA method uses the fact that the DM variations are correlated between adjacent observations, and applies a time-domain spectral analysis to measure the DM power-law spectrum and then extract the DM waveform by constructing a linear optimal filter \cite{lee14}.

\section{Multi-path scattering}

\begin{figure}
	\begin{center}
		\includegraphics[height=0.45\textwidth]{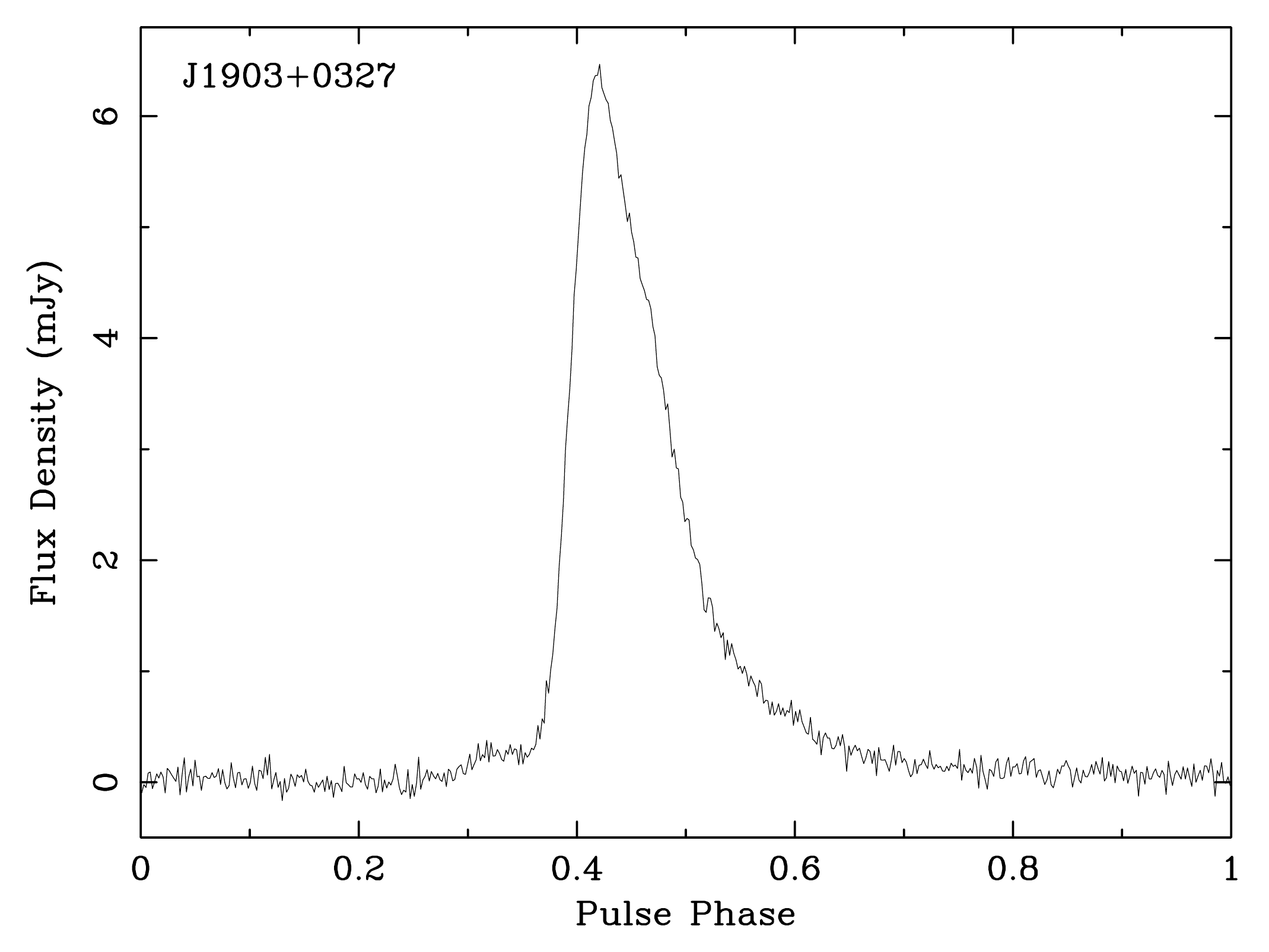}
	\caption{Pulse profile of PSR\,J1903+0327 at an observing frequency of 1.4\,GHz, showing an exponential scattering tail. This pulsar has a large DM\,=\,298\,pc\,cm$^{-3}$, contributing to strong scattering of the profile even at this relatively high observing frequency.}
	\label{fig:scattering}
	\end{center}
\end{figure}

Spatial inhomogeneities in the ISM give rise to scattering of the radio waves from pulsars. Such multi-path scattering manifests itself in several ways, including refractive and diffractive intensity scintillations as well as pulse broadening. 

Direct detection of pulse broadening is predominantly possible for pulsars with high DM values, i.e. for sources with more free electrons along their line of sight. Strong scattering cause large delays in the TOAs and hence high-DM pulsars are usually not included in PTAs. One exception is PSR\,J1903+0327, which because of its high flux density is useful for PTAs even though it has a large DM value. An observation of this pulsar can be seen in Fig. \ref{fig:scattering}, showing a large scattering tail in its folded profile. 
For pulsars with more modest scattering, the pulse broadening can be measured indirectly through analyses of the diffractive scintillation pattern of the pulsar, which will be described later in this section. 

The most commonly used model to describe the ISM consists of a thin disk of turbulent plasma midway between the pulsar and the observer \cite{sch68}. When the pulses travel through the disk, inhomogeneities in the plasma introduce phase perturbations to the signal. In combination with the different relative velocities of the pulsar, the ISM, and the Earth, this contributes to a diffraction pattern being observed. 
The perturbations are correlated over a scintillation bandwidth ($\Delta \nu_{\rm d}$), which is inversely proportional to the scattering delay ($\tau_{\rm d}$), and over a scintillation timescale ($\Delta t_{\rm d}$). 
Such diffractive scintillation effects were first observed in pulsars by Lyne \& Rickett \cite{lyn68}, and originate in interstellar density structure of scale size $\sim 10^6-10^8$\,cm. 
The ISM plasma is often considered to be composed as a Kolmogorov medium, which implies scaling of the scintillation parameters as $\Delta \nu_{\rm d} \propto \nu^{4.4}$, and $\Delta t_{\rm d} \propto \nu^{1.2}$ \cite{cor85}. The diffractive scintillation pattern for a pulsar can change drastically on short time scales of weeks to months \cite{hem08}. Similar to the case of DM variations, the relative velocities of the pulsar and the ISM give rise to the time variable scattering delays. 

Larger structures in the ISM, with a size scale $\sim 10^{10}-10^{12}$\,cm, give rise to refractive scintillation effects. 
These effects arise through wavefront curvature, by causing the angle of arrival to diverge from the straight line of sight, and are observed through flux density variations in pulsars \cite{ric90}.

\begin{figure}
	\begin{center}
		\includegraphics[height=0.6\textwidth]{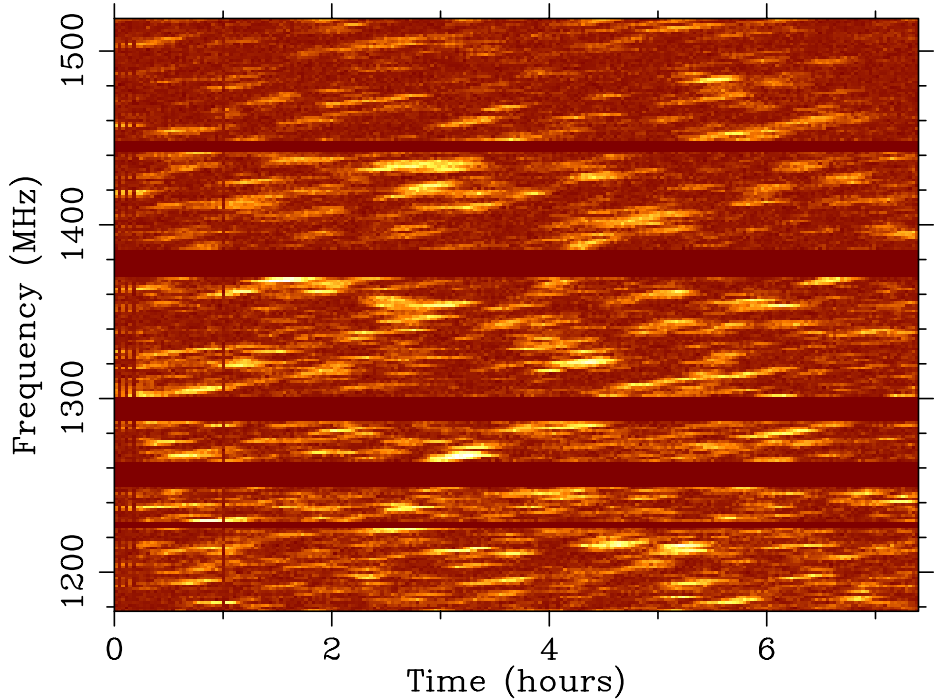}
	\caption{Dynamic spectrum of a $\sim$7-hr long observation of PSR\,J1614--2230, after interference signals have been removed from the data (showing up as dark lines in the plot). Lighter color indicates higher intensity of the pulsar signal, with separate bright islands referred to as scintles. Analyses of the scintles in the dynamic spectrum can be used to estimate scintillation parameters for a pulsar.}
	\label{fig:dynspec}
	\end{center}
\end{figure}

The multi-path broadening of pulses from diffraction give rise to the largest ISM delays, next to dispersive delays \cite{cor10}. 
These can be measured indirectly through dynamical spectrum analyses of the pulsar data. A dynamic spectrum displays the intensity of the on-pulse part of the pulse profile  over observing time and frequency. In the spectrum, scintillation maxima show up as bright islands on the map (so called ``scintles'', see Fig. \ref{fig:dynspec}). The scintles are limited in time since the interference pattern has a characteristic length scale, which becomes a characteristic time scale due to the relative velocity of the system, and they are limited in frequency because coherent radio beams have scattering delays of order $\tau_{\rm d}$.

To get enough signal to noise to measure the scintillation parameters from the scintles, a 2-dimensional autocorrelation function (ACF) is created from the dynamic spectrum. This 2D ACF is then summed in time and frequency separately, and the two resulting 1D ACFs are fitted with a Gaussian function, centered at zero lag. 
From the Gaussian fit, $\Delta t_{\rm d}$ is defined as the half-width at $e^{-1}$ of the summed frequency lag and $\Delta \nu_{\rm d}$ is the half-width at half-maximum of the summed time lag \cite{cor86}. 
The scattering delay, $\tau_{\rm d}$, can be calculated from $\Delta \nu_{\rm d}$ through
\begin{equation}
	2\pi \Delta \nu_{\rm d} \tau_{\rm d} = C_{\rm 1} 
	\label{eq:delay}
\end{equation}
where $C_{\rm 1}$ is a constant that depends on the geometry and spectrum of the electron density, and with expected value ranging from 0.6$-$1.5 \cite{lam99}. 

By creating dynamic spectra of each observation for each of the PTA pulsars, the scintillation parameters can be monitored, and in particular the variation of the scattering delays can be analyzed. This is important, since a constant delay would not cause any additional errors in timing precision, no matter its magnitude, but a highly variable delay would contribute largely to the timing noise. Therefore, the maximum variation of the delays need to be monitored, and can be compared to the individual TOA error to see if the scattering is one of the dominating sources of noise in the data. 
At the commonly used observing frequency band centered at $\sim$1.4\,GHz, for most PTA pulsars, the scattering delay variations are in general much smaller than the current TOA uncertainty \cite{col10,lev14}, and hence are currently not limiting the timing precision. With longer data spans and new developments in instrumentation, the timing precision will improve, and hence a detailed scattering analysis may become increasingly important in the future. 

The latest developments in ISM mitigation include a technique to estimate pulse broadening times and scattering delays using the signal-processing tool Cyclic Spectroscopy (CS; \cite{dem11,wal13}). This technique allows determination of the ISM impulse response function and extraction of the intrinsic pulse profile for highly scattered pulsars with a large flux density. CS backends have been developed at some of the radio telescopes \cite{jon14}, providing observations with much higher frequency resolution than possible with the filter banks used in conventional pulsar backends. Hence, very narrow scintillation bandwidths of high DM pulsars, that would not have been possible to measure with other pulsar backends, may be resolved in observations with a CS backend. This makes CS a very powerful tool for ISM mitigation and it will surely be used increasingly in the future.

\section{Conclusions}
PTAs strive to time MSPs with the highest possible precision, to allow for direct detection of nano-Hz GWs. 
Mitigating the delays that originate from radio wave propagation through the ISM is an essential part of the high-precision timing procedure. 

The largest ISM delays are due to dispersion, and time-variable DM values are currently corrected for in the timing data. Optimal DM correction is a highly topical issue among the PTA members and development of new techniques in this area is on-going. 

Further, interstellar density fluctuations also cause time-variable delays in the pulse time of arrival. For most PTA pulsars, scattering is currently not a limiting factor. However, with new wide-band receivers and higher timing accuracy, ability to correct for scattering delays may become an increasingly important part of high-precision timing. 
In addition, development of new techniques to properly mitigate scattering delays from the pulsar signal may allow for observations at lower frequencies as well as inclusion of more distant pulsars to the array in the future.

\section*{Acknowledgements}
The NANOGrav project receives support from the National Science Foundation (NSF) PIRE program award number 0968296.
The author would like to thank the NANOGrav collaboration for very useful advise and support, and for providing the pulsar data contributing to the figures in this document. 
She is also grateful to the organizers of the 10th International LISA Symposium for inviting her to present this talk.

\section*{References}
\bibliography{LISA}{}

\end{document}